\documentclass[aps,prb,showpacs,twocolumn]{revtex4}
\usepackage{epsfig}
\usepackage{amsmath}

\begin{document}

\title{Finite doping of a one-dimensional charge density wave: solitons vs. Luttinger 
liquid charge density}
\author{Yuval Weiss, Moshe Goldstein and Richard Berkovits}
\affiliation{The Minerva Center, Department of Physics, Bar-Ilan University,
  Ramat-Gan 52900, Israel}

\begin{abstract}
The effects of doping on a one-dimensional wire in a charge density wave state are studied using the 
density-matrix renormalization group method. We show that for a finite number of extra 
electrons the ground state becomes conducting but the particle density along the wire 
corresponds to a charge density wave with an incommensurate wave number determined by 
the filling. We find that the absence of the translational invariance can be discerned even
in the thermodynamic limit, as long as the number of doping electrons is finite.
Luttinger liquid behavior is reached only for a finite change in the electron filling factor,
which for an infinite wire corresponds to the addition of an infinite number of electrons.
In addition to the half filled insulating Mott state and the conducting states, we find
evidence for subgap states at fillings different from half filling by a single electron or hole.
Finally, we show that by coupling our system to a quantum dot, one can have a
discontinuous dependence of its population on the applied gate voltage in the 
thermodynamic limit, similarly to the one predicted
for a Luttinger liquid without umklapp processes.
\end{abstract}


\maketitle

\section{introduction}

The transition between a Mott insulator and a metallic 
phase in one-dimensional (1D) wires has attracted notable interest in recent 
years \cite{mott90}. 
Different types of transitions are possible, controlled by various physical parameters, 
such as the electron-electron interaction range and its strength. 
Since the Mott state exists only for commensurate fillings, 
the electron filling factor inside the wire is an important 
parameter, and it can be varied either by controlling the chemical potential, 
or by doping with a finite number of electrons. However, 
these two methods are different: doping with an infinitesimally small number of 
electrons breaks immediately the Mott insulator state, while due to the Mott gap, a finite change in the 
chemical potential is required in order to insert the first electron and 
cause the transition \cite{giamarchi03}. 

The metallic phase resulting from doping is usually described by the Tomonaga-Luttinger liquid (TLL)
theory \cite{voit94,giamarchi03}, although for a small finite doping, a large
curvature of the elementary excitation spectrum is expected, and deviations from the
TLL theory are possible.
In the TLL, many physical properties are determined by the
parameter $K$ which describes the interactions.
Recently, a discontinuity in the occupation
of a resonant level which is coupled to a TLL with $K<1/2$
was predicted by Furusaki and Matveev \cite{matveev}.
However, since the $K<1/2$ regime corresponds to very strong 
repulsive interactions, neither experimental nor numerical evidence
for such a jump has been obtained so far.

As a generic model for the wire it is convenient to use a tight-binding description of a 1D
lattice with spinless electrons. When repulsive interactions between nearest-neighbor electrons 
are considered, the Mott state, taking the form of a charge density wave (CDW),
occurs for strong enough interactions. The Hamiltonian of such a system can be written as
\begin{eqnarray} \label{eqn:H_wire}
{\hat H_{wire}} =
&-&t \displaystyle \sum_{j=1}^{L-1}({\hat c}^{\dagger}_{j}{\hat c}_{j+1} + H.c.) \\ \nonumber
&+&I \displaystyle \sum_{j=1}^{L-1}({\hat c}^{\dagger}_{j}{\hat c}_{j} - \frac{1}{2})
({\hat c}^{\dagger}_{j+1}{\hat c}_{j+1} - \frac{1}{2}),
\end{eqnarray}
where $I$ denotes the nearest-neighbor interaction strength, and the
hopping matrix element between nearest neighbors, $t$, sets the energy scale.
${\hat c}_j^{\dagger}$ (${\hat c}_j$) is the creation (annihilation)
operator of a spinless electron at site $j$ in the $L$-site wire, and
a positive background is included in the interaction term.

The model of Eq.~(\ref{eqn:H_wire}) is equivalent to that of the XXZ spin $1/2$ chain
by the Jordan-Wigner transformation\cite{jordan28}. The XXZ model is
exactly solvable using the Bethe ansatz \cite{bethe31,baxter89}. 
From this equivalence it is known that for a half filled system with periodic boundary conditions, 
a phase transition between a TLL phase and a CDW one occurs at $I=2t$. 
Of course, for sufficiently long wires the type of the boundary conditions does
not change this result \cite{our1,our2}. 
The interaction parameter $K$ is then given by \cite{g_formula}
\begin{eqnarray} \label{eqn:kappa}
{K} = \frac{\pi}{2 \cos ^{-1}(-I/2t)},
\end{eqnarray}
so that in the TLL phase, $I<2t$ and $K>1/2$, while
for $I>2t$, the half filled wire is no longer in the TLL phase, 
but is rather in a CDW state.

The competition between the TLL and the CDW phases is attributed to
the presence of umklapp processes in this model. For half filling the CDW phase
wins over the TLL phase once the interactions are strong enough. However,
when the wire is not exactly half filled, a CDW state cannot emerge since it demands
a commensurate filling, and thus the TLL description is valid even for interaction values 
which are greater than $2t$. As a result, for strong enough interactions, and 
sufficiently close to half filling, one can then get values of $K$ 
which are less than $1/2$. \cite{haldane}

Another approach to treat the Hamiltonian of Eq.~(\ref{eqn:H_wire})
is to map it, using bosonization \cite{giamarchi03}, into
that of the sine-Gordon model \cite{sg_model}.
In this model, the elementary excitations are solitons, which
carry half an electron charge \cite{giamarchi03}. 
In the vicinity of the Mott state the value of the TLL interaction parameter is
$K \rightarrow \frac{1}{4}^+$. The value $K = 1/4$, which cannot be obtained
using this model, corresponds to the Luther-Emery line, for which the solitons 
are effectively non interacting \cite{luther74}. For small deviations from half filling
the solitons are only weakly interacting.

In this paper we investigate the effects of a finite doping on a 1D Mott state. We 
show that although a small number of electrons results in an insulator-metal transition,
the charge distribution in the metallic system with a finite number of electrons is 
not uniform, but rather corresponds to an incommensurate CDW.
The added charge can also be considered as delocalized solitons in the original 
commensurate CDW, whose spatial distribution is determined by boundary
conditions and symmetry considerations. Nevertheless, the system is conducting, and once a bias 
voltage is applied the solitons are free to move and transport charge across the wire.

An exceptional behavior is exhibited when a lattice with an even number of sites $L=2p$
is occupied by $p \pm 1$ electrons, i.e., there is a single additional electron (or hole) relative to half filling.
As will be shown, in this case the insulating behavior of the original Mott state 
is retained even though the filling is incommensurate. 
This unusual case will be explained as a result of subgap states in the soliton energy spectrum.

In addition, we show that since the Mott state with a small doping allows for the interaction 
parameter in the range $1/4<K<1/2$, it might exhibit
the Furusaki-Matveev discontinuity. The appearance of that jump is not a foreclosed
conclusion, since the model used by Furusaki and Matveev does not contain
umklapp processes. Nevertheless, we find that the jump indeed occurs, and show its implications 
on the solitons present in the wire near half filling.

Since the CDW state is obtained for exactly half filling, it is convenient to denote 
the extra charge by $Q=n_e-L/2$, where $n_e$ is the number of electrons in the system. 
As will be shown, the parity of $L$ plays a significant role in the 
behavior when adding the first few electrons.
Using the finite-size version of the density-matrix renormalization group (DMRG) 
method \cite{white93,schollwock05} the Hamiltonian $\hat H_{wire}$ was diagonalized and
the ground state was calculated for different values of the charge $Q$ and the lattice size $L$.
In the current paper we present results obtained for $I=3t$, which is deep inside 
the CDW regime (for half filling). Nevertheless, other interaction strengths in the 
regime of $I>2t$ have been checked as well, and were found to lead to qualitatively similar results.

The outline of the paper is as follows: in the next section, Sec. \ref{sec:cdw}, we present
the charge occupation along nearly half filled wires, which demonstrates the 
presence of incommensurate CDW or solitons.
In Sec. \ref{sec:MIT} we discuss the excitation spectrum of the solitons, and the
addition spectrum of the system. 
For systems which are exactly half filled, or doped with only one electron or hole, we show 
that the ground state is insulating; otherwise, it is conducting.
In Sec. \ref{sec:FMJ} we
discuss the existence of the Furusaki-Matveev jump when the doped wire is coupled to
a resonant level. Finally we conclude in Sec. \ref{sec:conc}.

\section{Charge distribution}
\label{sec:cdw}
\begin{figure}[htbp]
\centering
\begin{minipage}[t]{.45\textwidth}
\includegraphics[trim=0mm 0mm 0mm 0mm, clip, width=.95\textwidth]{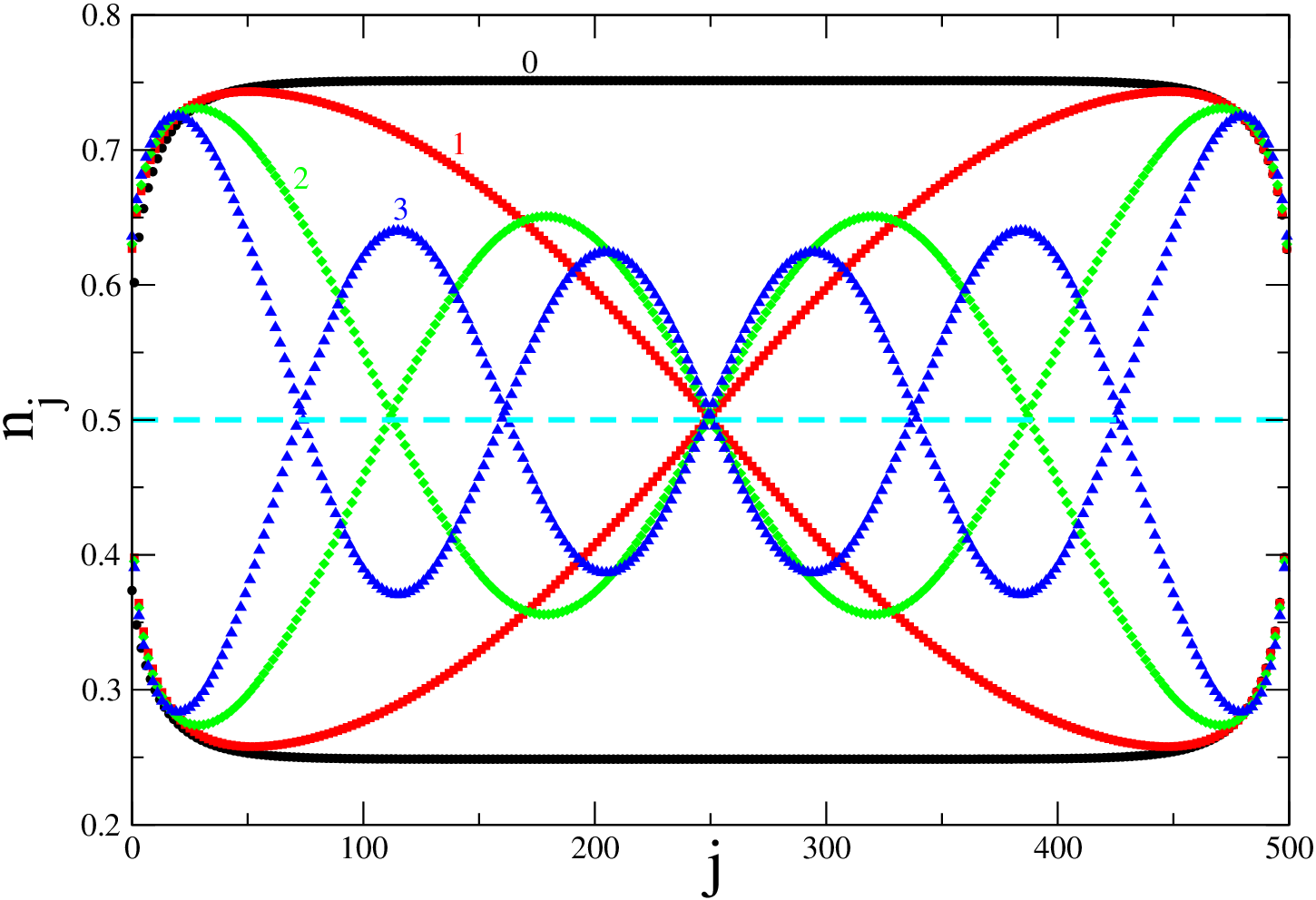}
\end{minipage}
\vskip 1truecm
\begin{minipage}[t]{.45\textwidth}
\includegraphics[trim=0mm 0mm 0mm 0mm, clip, width=.95\textwidth]{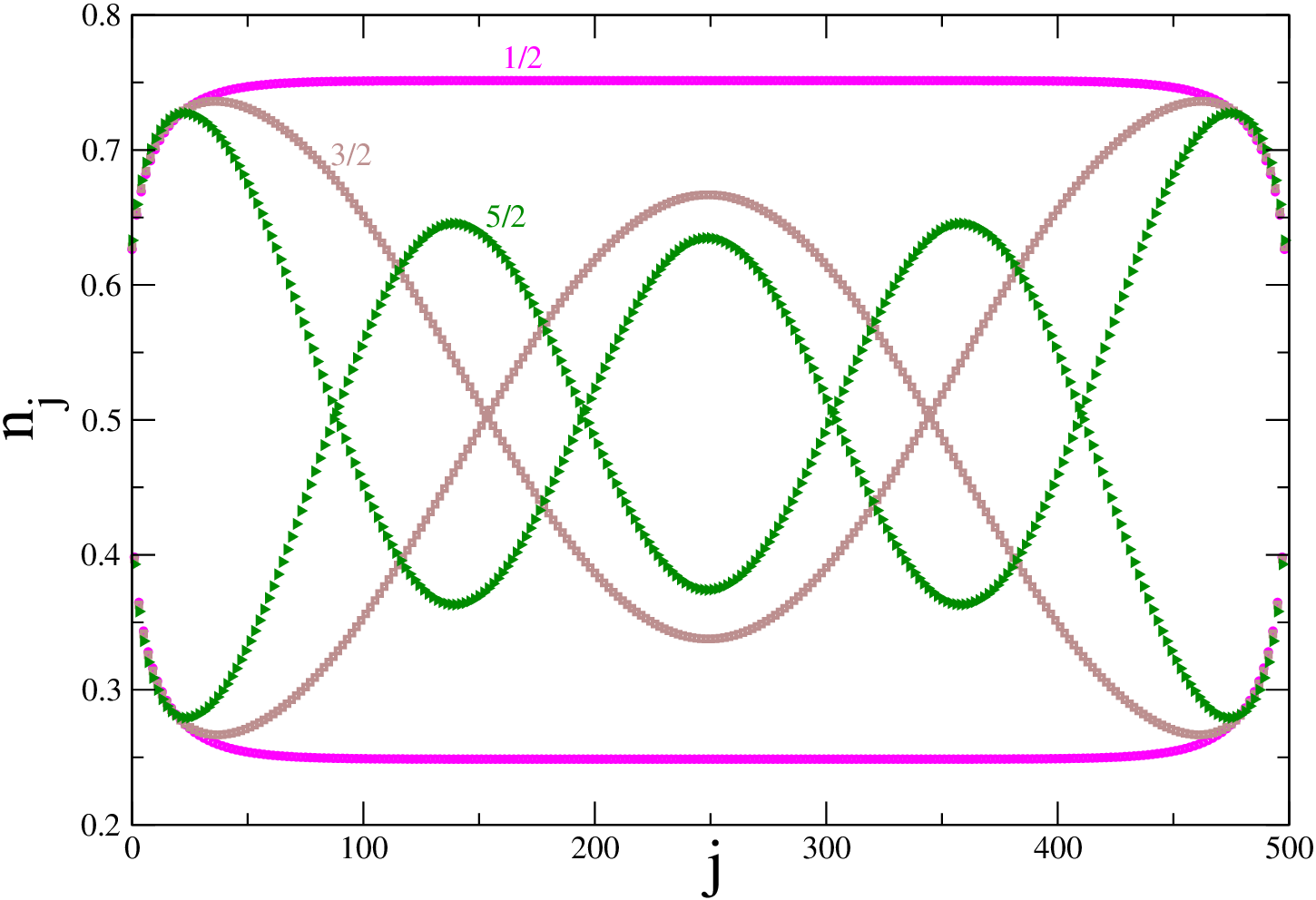}
\end{minipage}
\caption[Solitons in a $500$-sites charge density wave]
{\label{fig:L500_sol}
(Color online) The electrons occupation in a lattice of $500$ sites (upper panel) and $499$ sites
(lower panel) for different number
of electrons. The values near the curves denote the extra charge $Q$ 
which varies between $Q=0$ for half filling and $Q=3$. 
}
\end{figure}

We begin by presenting the occupation of electrons along the wire for different numbers of
extra electrons $Q$ for wires with an odd or an even number of sites 
(Fig.~\ref{fig:L500_sol}). The $Q=0$ case (for an even $L$) 
results in a flat distribution of $n_j=1/2$ for each lattice site $j$ (dashed line). 
This state is a linear combination of two degenerate CDW states, which are 
eigenstates of the Hamiltonian in the thermodynamic limit. 
These CDW states, which are also shown in Fig.~\ref{fig:L500_sol}, are numerically obtained 
by applying an infinitesimally small potential on the first site, which
breaks the degeneracy between them\cite{our1}.
The population of such a CDW state along the wire (near the center) can be written as
\begin{eqnarray} \label{eqn:n_j_cdw}
{n_j} = n + A \cos(2 \pi n j + \phi),
\end{eqnarray}
where the amplitude $A$ depends on the interaction strength, $n=1/2$ is the filling, and $\phi$ is a phase.

When $Q \ne 0$, the ground state is no longer degenerate and the electron
occupation throughout the lead is not uniform. Let us start with the $Q=1/2$ case 
(when $L$ is odd), which results in a true CDW state, i.e., for each even (odd)
site the occupation is low (high). Unlike the case of the degenerate ground states for $Q=0$, 
the $Q=1/2$ ground state is not coupled by the Hamiltonian to the similar CDW ground state of 
$Q=-1/2$ due to the difference in their total population. 

\begin{figure}[htbp]
\vskip 1truecm
\centering
\includegraphics[width=3in,height=!]{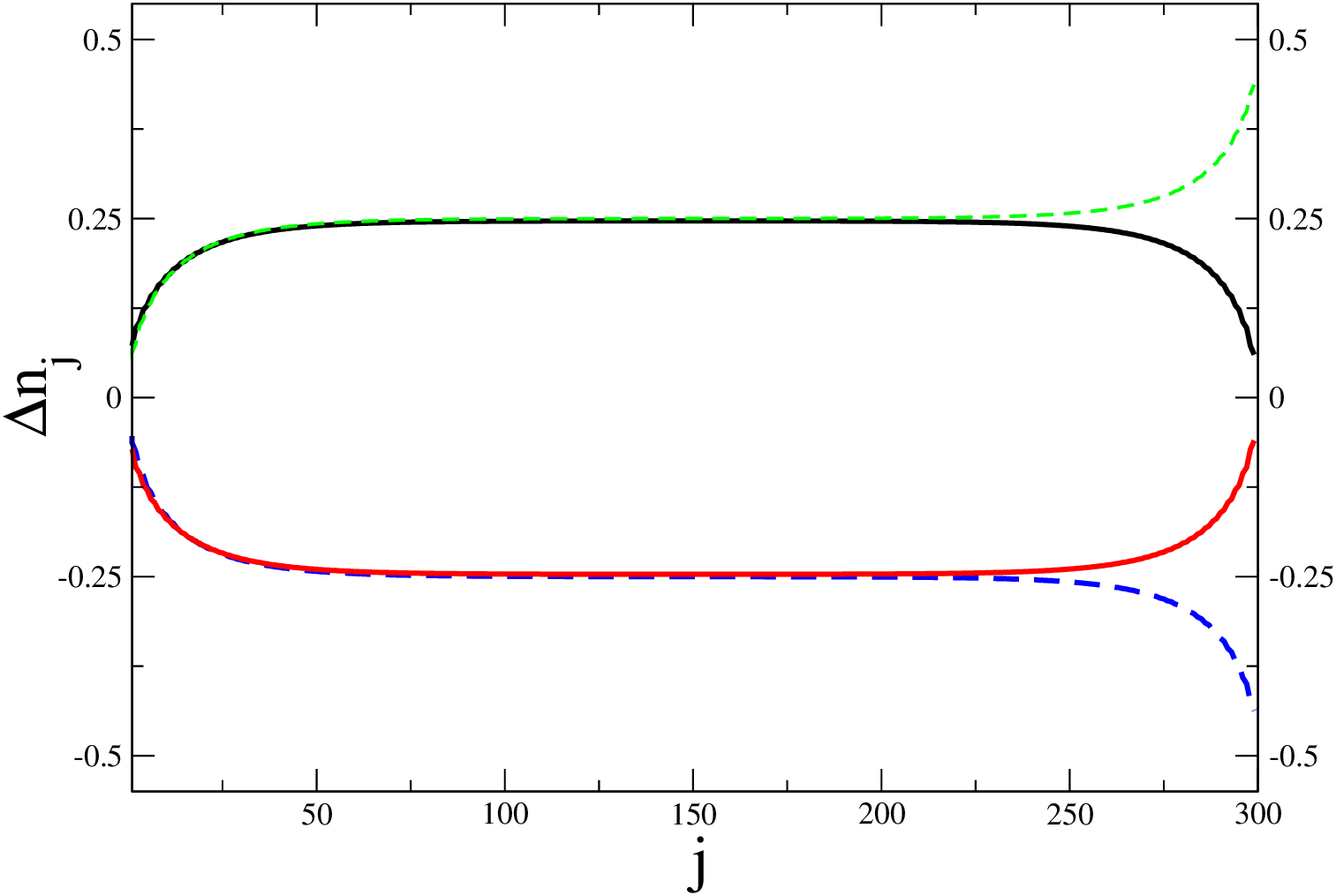}
\caption[]
{\label{fig:DN_j}
(Color online) The extra charge distribution along a wire of length $L=300$ sites with $Q=0$ (solid lines)
and of $L=299$ sites with $Q=\pm 1/2$ (dashed lines).
}
\end{figure}

In order to explore the addition of electrons onto the CDW states,
one can take the uniform distribution of the $Q=0$ state as a reference, 
and investigate the difference from it by calculating the accumulated population
$\Delta n_j = \sum_{i \le j}n_i-j/2$. 
As can be seen in Fig.~\ref{fig:DN_j},
$\Delta n_j$ for each of the two clean CDW states of $Q=0$ (presented in solid lines), 
has an extra charge of $e/4$ localized at one of the wire edges,
and a compensating charge (i.e., $-e/4$) localized at the other. The difference
between these two states is thus a soliton (of charge $e/2$) located at one edge of the wire
and an antisoliton (of charge $-e/2$) at the other edge.

By taking one of the $Q=0$ CDW states, and 
locating a soliton at its negatively charged edge, while leaving the other edge as it is,
one gets a new CDW state in which both edges have a positive charge, i.e., a
CDW state having $Q=1/2$. Similarly, placing an antisoliton at the edge with the positive 
charge of the $Q=0$ state results in the $Q=-1/2$ CDW state. These states are also shown
in Fig.~\ref{fig:DN_j} (dashed lines). 

For the $Q=1$ case (when $L$ is even), adding a localized soliton
near one of the edges is not sufficient, and it requires the addition of another 
charge of $e/2$. This extra charge is obtained by the formation of
a delocalized soliton, centered at the middle of the wire.
Further increase of the filling above $Q=1/2$ ($Q=1$) for odd (even) wire lengths,
results in the addition of two delocalized solitons for every electron,
so that the total number of delocalized solitons is $2Q-1$.

\begin{figure}[htbp]
\vskip 1truecm
\centering
\includegraphics[width=3in,height=!]{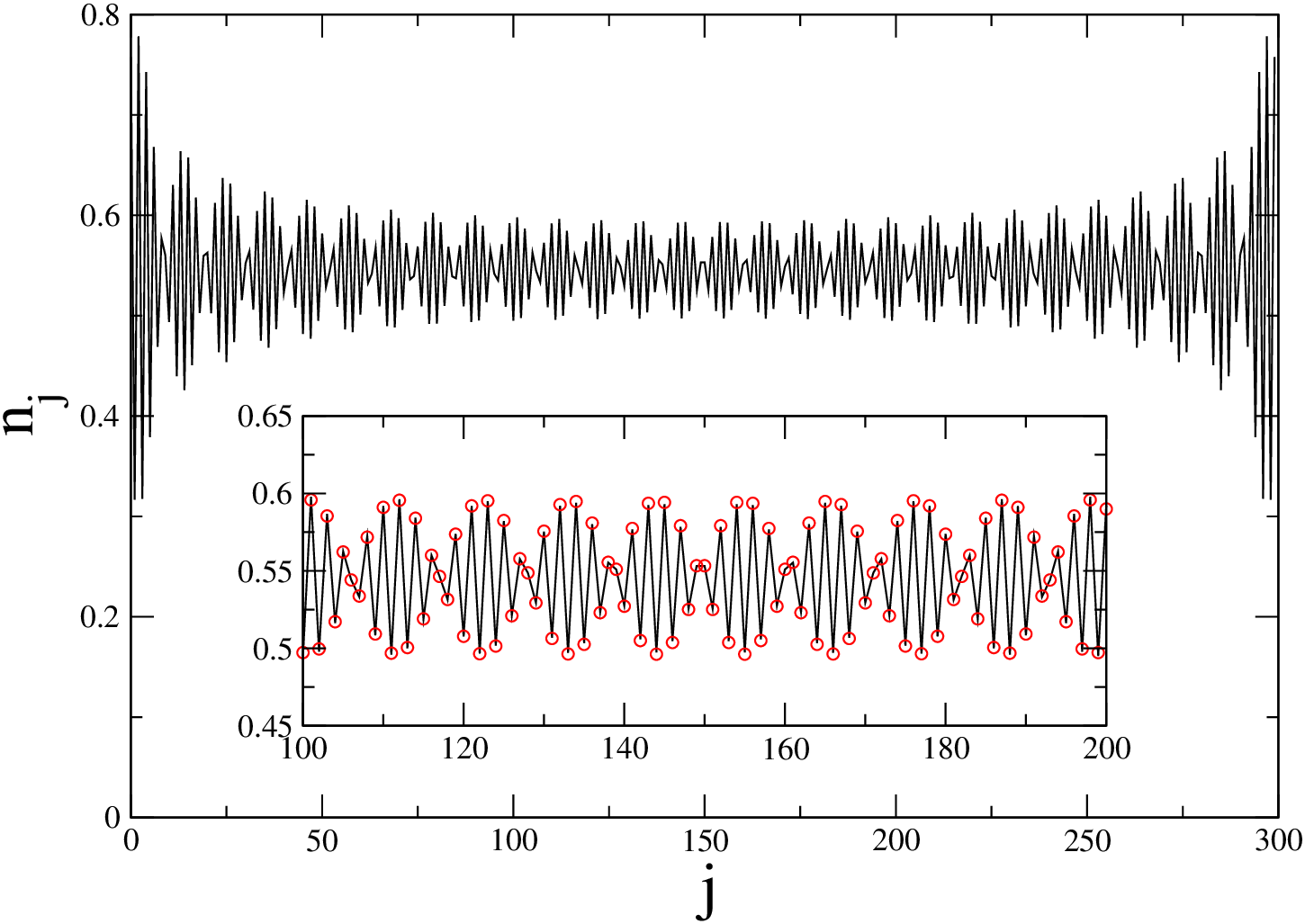}
\caption[Solitons in a $300$-sites charge density wave with $27$ additional electrons]
{\label{fig:L300_n14}
(Color online) The electrons occupation in a lattice of $300$ sites with additional $14$ electrons.
A fit to Eq.~(\ref{eqn:n_j_cdw}) (done over the middle region) is shown in the inset 
(symbols).
}
\end{figure}

The delocalized solitons, which carry a charge of $e/2$ each, are free to move along the 
wire. However, when no bias voltage is applied, the charge distribution across the wire is
fixed by the boundaries and by symmetry considerations. Practically these constraints lead 
to a density wave with a cosine term similar to that of Eq.~(\ref{eqn:n_j_cdw}), 
in which the filling factor $n$ is modified. 
This statement is true for much larger values of $Q$ as well. 
For instance, Fig.~\ref{fig:L300_n14} shows
the electron distribution for a $300$-site wire
with $Q=14$, i.e., with $27$ delocalized solitons. It is easy to see that although
the amplitude of the oscillations is reduced, it doesn't vanish and it is rather constant near the middle
of the wire. Moreover, the oscillations can still be fitted to Eq.~(\ref{eqn:n_j_cdw})
using $n$ as a fitting parameter. As can be expected, this results in
a value of $n$ which matches the filling of the wire.

\begin{figure}[htbp]
\centering
\includegraphics[width=3in,height=!]{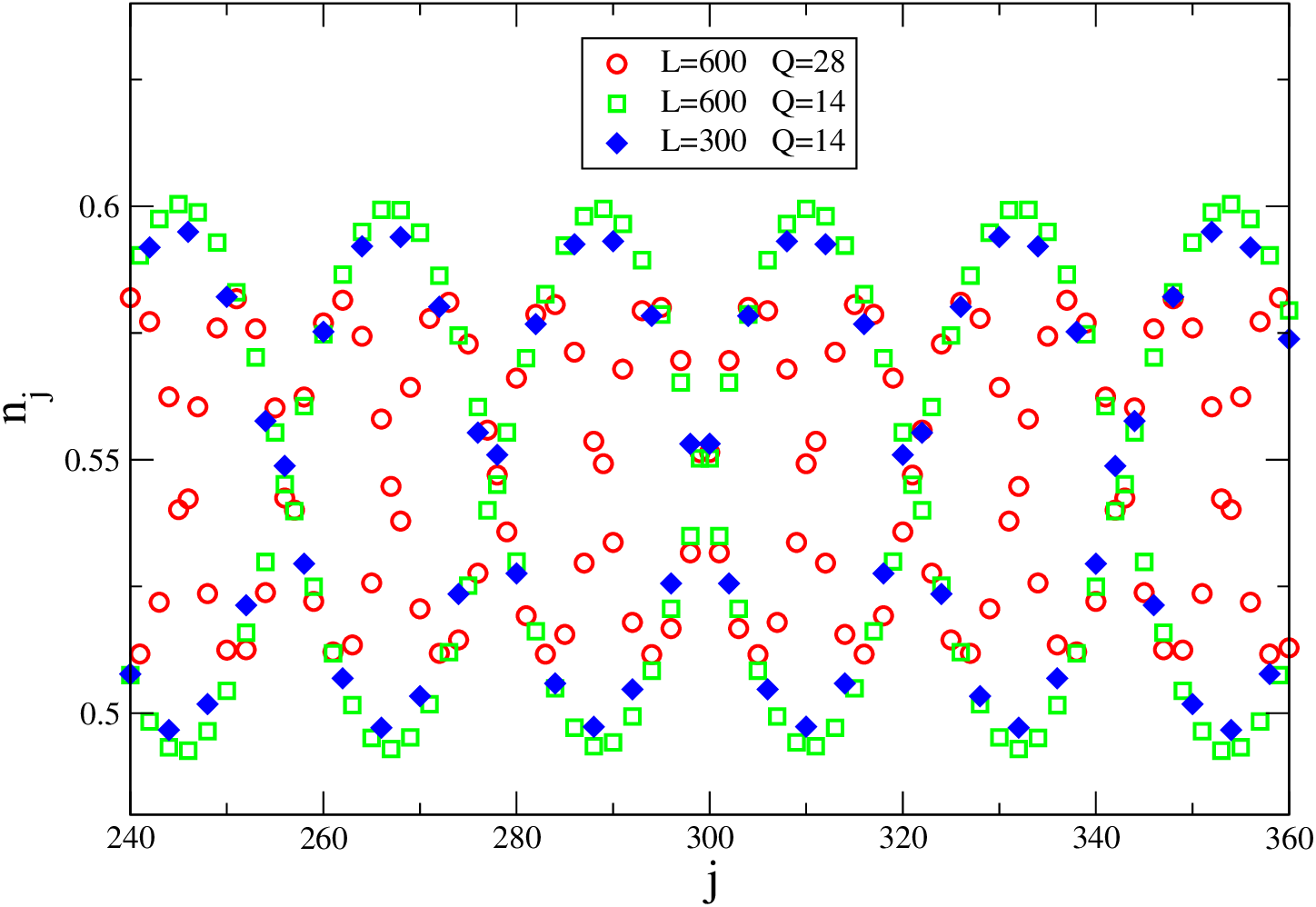}
\caption[Solitons survive in the thermodynamic limit]
{\label{fig:L300_600_n14_28}
(Color online) The electrons occupation near the middle of the system in lattices of 
$300$ or $600$ sites with additional $14$ 
or $28$ electrons. Results for the $300$-site wire are shown as a function of $2j$.
}
\end{figure}

The non-uniform charge distribution for a fixed number of electrons
is not a finite-size effect,
and survives in the thermodynamic limit as well.
In Fig.~\ref{fig:L300_600_n14_28}
one can see a comparison between wires of $300$ and $600$ sites, with
$Q=14$ and $Q=28$. One can see weaker oscillations in the $Q=28$ case, 
attributed to a smaller wave length. 
On the other hand, increasing the size of the lead from $L=300$ to $L=600$ while maintaining 
the number of extra electrons $Q$ constant, results in an increase of the oscillations.

One can thus conclude that on the one hand the oscillations are expected to vanish 
if a finite filling fraction is considered, i.e., in the limit of 
$Q \rightarrow \infty$, $L \rightarrow \infty$, keeping $Q/L$ constant.
On the other hand, when a finite number of electrons is added, 
the oscillations are expected to be noticed even in the limit of 
$L \rightarrow \infty$.

\section{Excitation spectrum}
\label{sec:MIT}
Up to now we have shown that the density distribution of the electrons along the
wire, when it is doped by a finite number of electrons, is not flat, and preserves
some features of the Mott state. One can still wonder whether these doped states 
retain the insulating behavior of the Mott state or not.
To clarify that point we study the size dependence of the addition spectrum, defined through 
\begin{eqnarray} \label{eqn:Delta2}
\Delta_2(Q) = E_0(Q+1)-2E_0(Q)+E_0(Q-1),
\end{eqnarray}
where $E_0(Q)$ is the ground-state energy of the wire with $Q$ electrons above half filling.
For a Mott state the limit of $\Delta_2$, as $L \rightarrow \infty$,
should be a finite value, corresponding to the gap size, while for a conducting state 
one expects $\Delta_2 \rightarrow 0$.

\begin{figure}[htbp]
\centering
\includegraphics[width=3in,height=!]{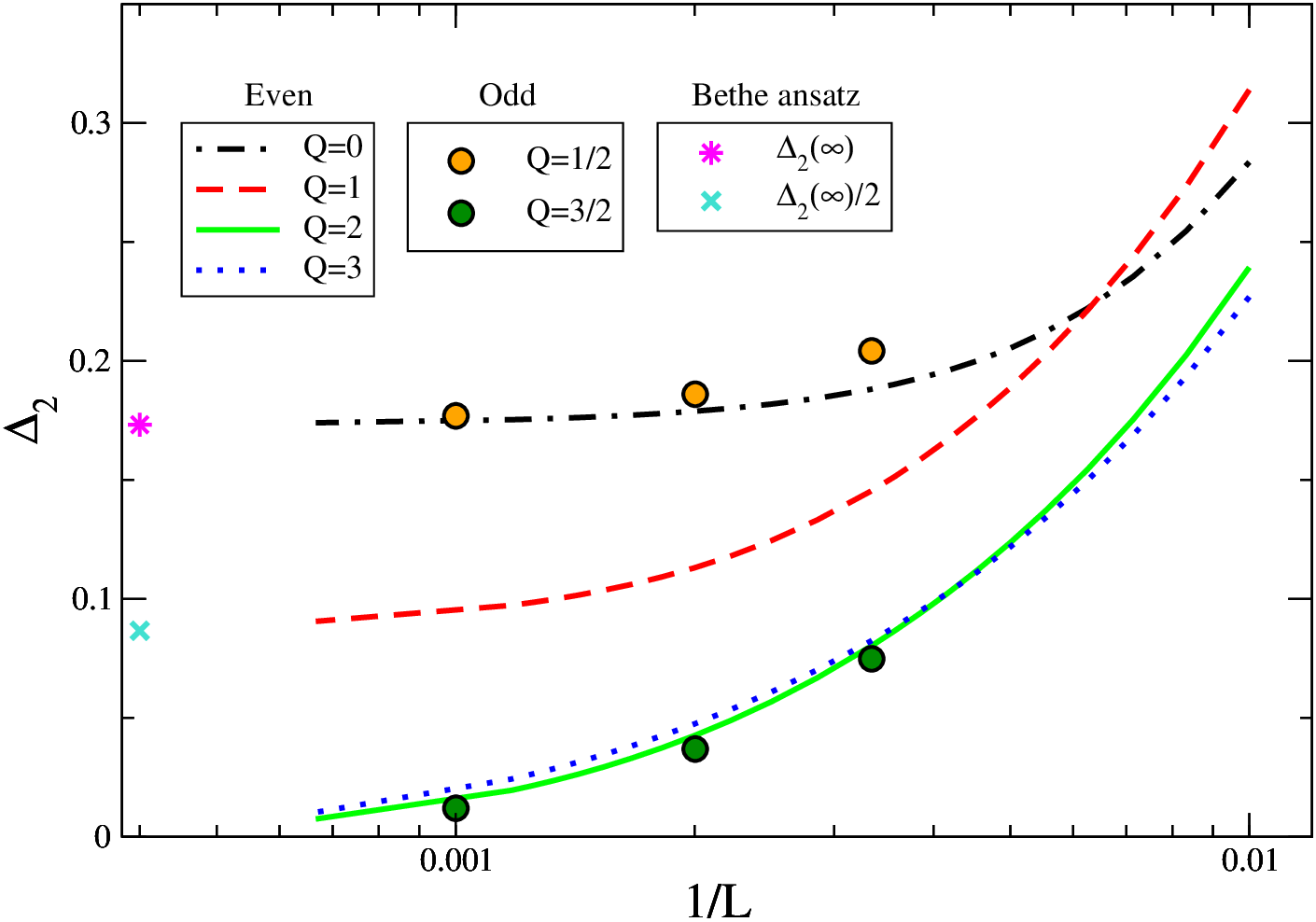}
\caption[Addition spectrum for different fillings]
{\label{fig:MIT}
(Color online) The dependence of the addition spectrum on the lattice size $L$, for different 
fillings, comparing the cases of even and odd sizes and the exact Bethe
ansatz results. Note the semi-logarithmic scale.
}
\end{figure}

The dependence of $\Delta_2(Q)$ on the wire length $L$ is presented in Fig.~\ref{fig:MIT}
for even wire sizes (lines) and for odd sizes (filled symbols), for different
number of electrons $Q$. As can be seen, for $Q=0$ and $1/2$, the values of $\Delta_2$
converge to the value $\Delta_2(\infty)$ (presented by the asterisk symbol at
the left margin of the figure), given by the Bethe ansatz \cite{bethe31,baxter89}. 
On the other hand, for $Q \ge 3/2$ one gets $\Delta_2 \rightarrow 0$, which indicates 
that these states are conducting. 

A deviation from this intuitive picture appears for $Q=1$,
in which an unexpected gap of $\Delta_2(\infty)/2$ occurs. Nevertheless, this result can
be explained by the spectrum of soliton states near half filling. It is known that for
solitons in the sine-Gordon model with open boundary conditions 
there are two degenerate subgap states at zero energy, positioned
exactly between the conduction and the valence bands, each of them localized near
one of the edges of the system \cite{gogolin}. In our case these subgap 
states are thus the localized solitons, which we identified as the difference 
between the two CDW states of $Q=0$ presented in Fig.~\ref{fig:L500_sol}.

\begin{figure}[htbp]
\vskip 1truecm
\centering
\includegraphics[width=3in,height=!]{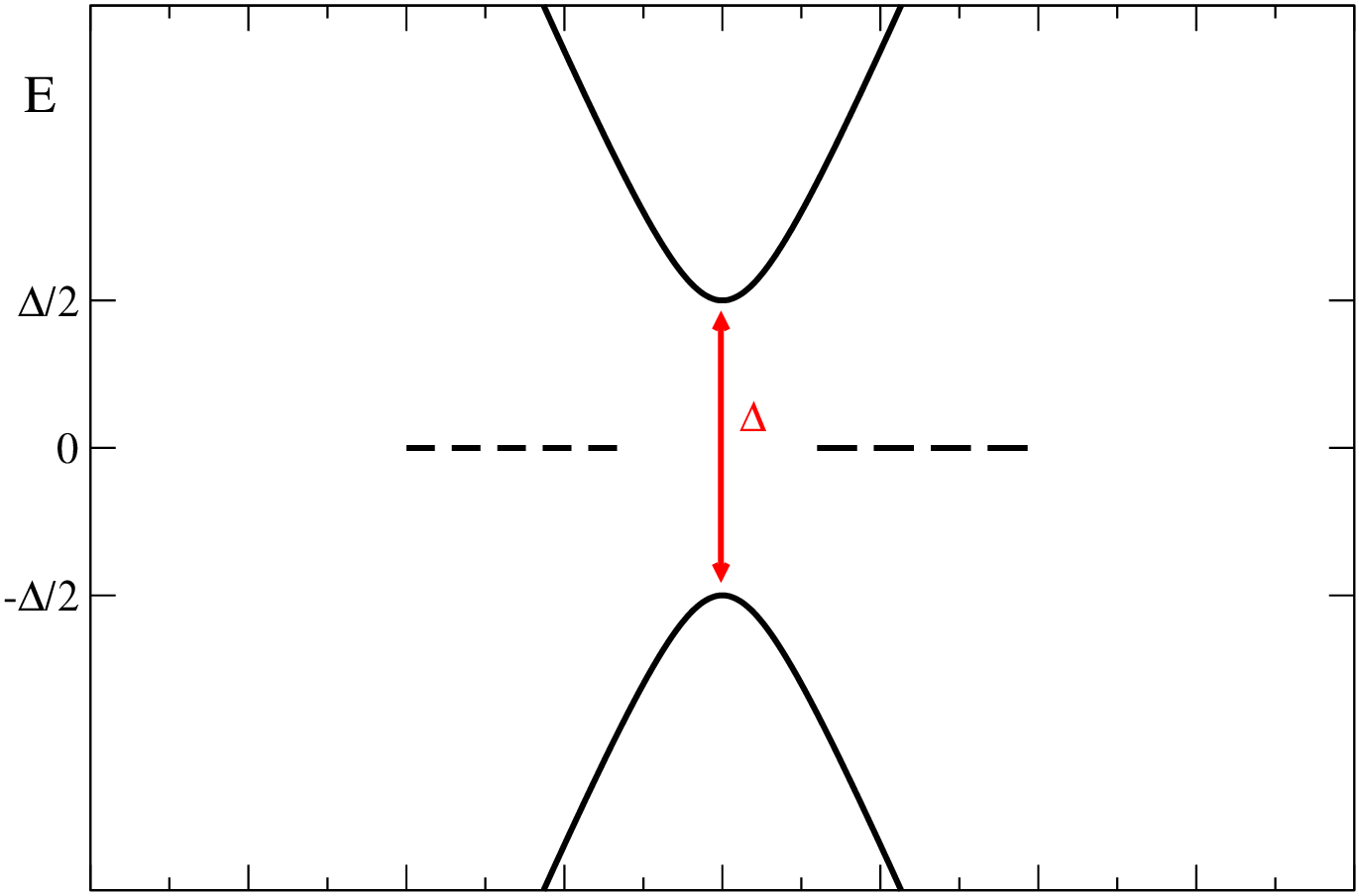}
\caption[Energy bands of the solitons]
{\label{fig:Ebands}
(Color online) A schematic picture of the energy bands of the solitons according to 
Ref.~\onlinecite{gogolin}. The bands are separated by energy $\Delta$,
and two discrete states exist at zero energy.
}
\end{figure}

Thus the energy spectrum of the solitons is represented by valence and conduction bands
separated by a gap $\Delta$, while the subgap states exist at $\Delta/2$ above the valence 
band (see schematic picture in Fig.~\ref{fig:Ebands}). 
The filling of the subgap states with the localized solitons does not change the total energy. 
However, every additional soliton increases the energy in $\Delta/2$ (in the limit $L \rightarrow \infty$). 

For an odd $L$, the states with $Q=\pm 1/2$, which have the same energy due to the
particle-hole symmetry of the Hamiltonian, differ only in the occupation of the two 
localized solitons. The addition of any other electron is equivalent to the addition 
of two delocalized solitons, so that the energy difference between successive fillings is $\Delta$. 
Therefore, $E_0(1/2) - E_0(-1/2)=0$ and $E_0(3/2) - E_0(1/2)=\Delta$,
resulting in $\Delta_2(1/2)=\Delta$. For higher values of $Q$, one gets
$E_0(Q+1) - E_0(Q)=\Delta$, so that $\Delta_2(Q \ge 3/2)=0$.
These results are summarized in Table~\ref{tbl:sol_eg_odd}.

\begin{table}[ht]
\centering
\begin{tabular}{|c||c|c|c|c|c|}
\hline
$Q$ & $-\frac{1}{2}$ & ~$\frac{1}{2}$~ & ~$\frac{3}{2}$~ & $\frac{5}{2}$ & $Q^\prime>\frac{5}{2}$ \\ [1ex]
\hline

\hline
$E_0$ & 0 & 0 & $\Delta$ & $2\Delta$ & $(Q^\prime-\frac{1}{2})\Delta$ \\ [1ex]
\hline
$\Delta_2$ & $\Delta$ & $\Delta$ & 0 & 0 & 0 \\ [1ex]
\hline
\end{tabular}
\caption{\label{tbl:sol_eg_odd} Addition spectrum for wires with an odd number of sites
in the limit $L \rightarrow \infty$.}
\end{table}

On the other hand, when $L$ is even, the $Q=0$ state corresponds to the occupation of only
one localized soliton (or, more precisely, to a linear combination of two states, in each of them
one edge soliton state is filled and the other is empty). 
The addition of the first extra electron fills the additional localized 
soliton state and a single delocalized soliton state, so that $E_0(1)-E_0(0)=\Delta/2$.
Every additional electron adds two delocalized solitons, thus for $Q>1$ one gets 
$E_0(Q+1)-E_0(Q)=\Delta$. Therefore, as one can see in  Table~\ref{tbl:sol_eg_even},
$\Delta_2(0)=\Delta$, $\Delta_2(1)=\Delta/2$, and
$\Delta_2(Q \ge 2)=0$. 

\begin{table}[ht]
\centering
\begin{tabular}{|c||c|c|c|c|c|c|}
\hline
$Q$ & $-1$ & ~$0$~ & ~$1$~ & $2$ & $3$ & $Q^\prime>3$ \\ [1ex]
\hline

\hline
$E_0$ & $\frac{\Delta}{2}$ & 0 & $\frac{\Delta}{2}$ & $\frac{3\Delta}{2}$ & $\frac{5\Delta}{2}$ & $(Q^\prime-\frac{1}{2})\Delta$ \\ [1ex]
\hline
$\Delta_2$ & $\frac{\Delta}{2}$ & $\Delta$ & $\frac{\Delta}{2}$ & 0 & 0 & 0\\ [1ex]
\hline
\end{tabular}
\caption{\label{tbl:sol_eg_even} Addition spectrum for wires with an even number of sites
in the limit $L \rightarrow \infty$.}
\end{table}

Before commencing with the observation of Furusaki-Matveev jump 
we briefly summarize the results so far. We have demonstrated the existence of 
CDW states for $Q=0$ and $Q=\pm 1/2$, and shown that these states are insulating. The ground states with 
$Q=\pm 1$ were found to be insulating as well, with an excitation gap which is half of the Mott gap.
States with $Q>1$ are conducting and thus should be generally described by the TLL theory.
Nevertheless, we have found that the spatial distribution of the electron density
is not uniform as in the regular TLL picture. 
Furthermore, this distribution can be fitted to an incommensurate CDW form. 
It is therefore interesting to explore some other predictions of the TLL theory
for such states.

\section{Furusaki-Matveev jump}
\label{sec:FMJ}
The unconventional behavior of the gapless states with the non-uniform density can be 
illustrated by studying the 
discontinuity in the occupation of a resonant level coupled to the wire \cite{matveev}. 
As mentioned above, once $K < 1/2$, the level is expected to show a jump in its occupation
as a function of its energy.
Nevertheless, one should note that umklapp processes, 
which are an essential ingredient in explaining the behavior of our system, were not 
considered in the theoretical framework of Ref.~\onlinecite{matveev}. 

Furthermore, since here the particle distribution in the uncoupled wire is not flat, 
the occurrence of the Furusaki-Matveev discontinuity raises an additional question.
If the electrons occupation profile along the uncoupled wire was uniform, one 
should have observed Friedel oscillations in the wire as soon as it is coupled to the
resonant level. The potential of the resonant level can then be continuously 
changed until the predicted jump in the level occupation occurs. At that point,
an inversion of the Friedel oscillations along the wire is expected. 
On the other hand, once the wire is a Mott state doped by a finite number of electrons
and the electron density is not uniform, the situation
is less clear.

We first calculate the value of $K$ for the doped Mott state, and show that it is below $1/2$.
In order to have the 
same value of $K$ for different lengths of wires, one must consider an equal density 
of additional electrons, and we choose to work with doping of $Q=L/50$, with wires of sizes $100$ sites and above. 
$K$ may be obtained \cite{giamarchi03,schulz90} by the ratio between $\Delta E$, the energy difference 
between the ground state and the first excited state, and the addition spectrum $\Delta_2$,
since for spinless electrons with open boundary conditions $\Delta E=\frac{\pi v_c}{L}$ and $\Delta_2=\frac{\pi v_c}{KL}$. 
More accurate result may be obtained by fitting both $\Delta_2(L)$ and $\Delta E(L)$
to a polynom in $1/L$, and then obtaining $K$ from the ratio of the linear coefficients \cite{ejima05}.
Using both methods we find that the value of $K$ for our wire is $0.42$.

In order to represent the coupling of a resonant level or an impurity 
of energy $\epsilon_0$ to the left edge of the wire
the following term is added to the Hamiltonian:
\begin{eqnarray} \label{eqn:H_imp}
{\hat H_{\rm imp}} = \epsilon_0 {\hat a}^{\dagger}{\hat a}
-V_0 ({\hat a}^{\dagger}{\hat c_1 + H.c.)},
\end{eqnarray}
where ${\hat a}^{\dagger}~({\hat a})$ is a creation (annihilation) operator of an electron
in the resonant level, and $V_0$ is the hopping matrix element between the level
and the first site of the wire. Interaction between the impurity and the wire
is not considered here since it does not change the result qualitatively, but only causes a 
renormalization of the hopping amplitude $V_0$ towards larger values 
\cite{matveev,our1}. 

\begin{figure}[htbp]
\centering
\includegraphics[width=2.85in,height=!]{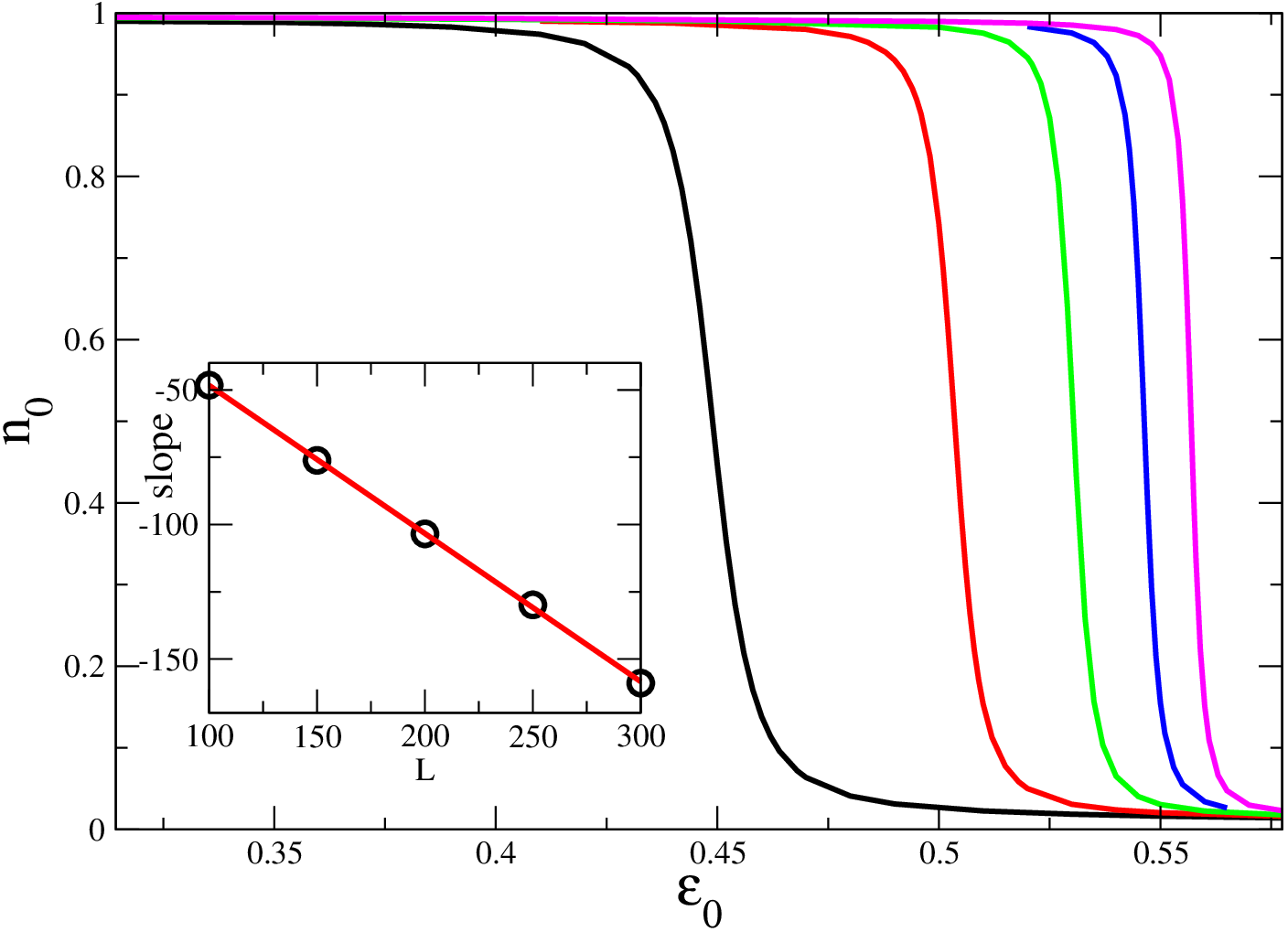}
\caption[]
{\label{fig:FMJ_slope_L}
(Color online) The occupation of an impurity which is coupled to one end of 1D lattices
of different sizes. The results shown are for lattice sizes between $100$ and $300$ (from left to right)
in steps of $50$. In order to compare cases with identical values of $K$, 
the number of additional electrons is taken as $Q=L/50$, so that $K=0.42$.
Inset: the slope near the point where $n_0=1/2$, which shows a linear dependence on the system size.
}
\end{figure}

The level occupation $n_0$ as a function of the level energy $\epsilon_0$ is presented in 
Fig.~\ref{fig:FMJ_slope_L} for different wire lengths using $V_0=0.2t$.
As mentioned above, wires in different lengths contain different number of additional 
electrons using $Q=L/50$, and $K=0.42$.
When $\epsilon_0$ is much larger than the chemical potential in the wire $\mu$, 
the impurity is almost empty, and the wire contains most of the $Q$ extra 
electrons. On the other hand, for $\epsilon_0 \ll \mu$ the impurity is almost entirely occupied, 
and then only $Q-1$ extra electrons are in the wire.

Although the population of the impurity is found to be continuous for all finite 
wire lengths studied, it is expected to display an abrupt jump in the thermodynamic limit, 
$L \rightarrow \infty$. In order to demonstrate this we
study the dependence of the occupation slope near $n_0=1/2$ on the wire length $L$.
As is clearly seen in the inset of Fig.~\ref{fig:FMJ_slope_L}, the slope scales 
linearly with $L$, which is a clear sign of a first order transition in the
thermodynamic limit \cite{binder84}. 

\begin{figure}[htbp]
\vskip 1truecm
\centering
\includegraphics[width=3in,height=3in]{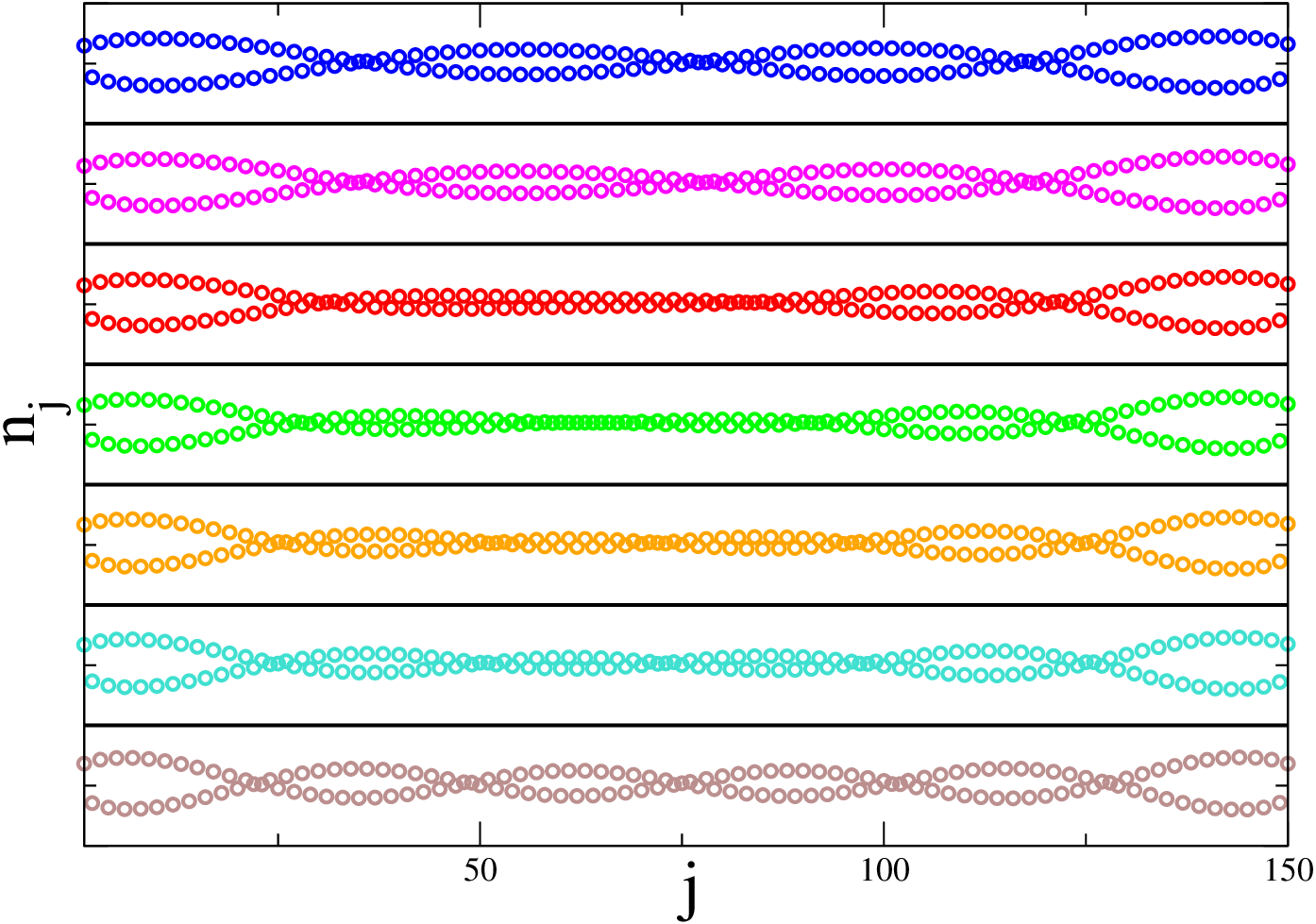}
\caption[]
{\label{fig:FMj_transition}
(Color online) The transition of an electron from the resonant level into the wire, which leads to
the formation of $2$ additional solitons. The results shown are
for a level coupled to a $150$-site wire, with $Q=3$. Different curves correspond to
values of $\epsilon_0$ between $0.498$ (top) to $0.510$ (bottom) with offsets 
in the vertical axis for clarity. 
}
\end{figure}

The finite size transition region, in which the electron transfers from the resonant level 
into the wire, is interesting by itself, since the addition of a single 
electron to the wire is related to the appearance of 
two additional delocalized solitons in the electron population inside it.
As can be seen in Fig.~\ref{fig:FMj_transition},
as $\epsilon_0$ increases the electron bound to the impurity tunnels
into the wire and additional two solitons are created in the wire. 
As mentioned above, this is in sharp
contrast to the case of a resonant level coupled to a TLL having a uniform charge distribution, 
in which only a local change of the Friedel oscillations in the charge distribution of the wire is expected.

\section{Conclusions}
\label{sec:conc}
In this paper we have shown that doping a Mott state with a finite number of
electrons yields states which preserve the charge modulations 
of the density wave even for an infinite system size. The charge carriers are solitons, 
whose spectrum fits the known predictions of the sine-Gordon model. 

We have demonstrated how the solitons are added to the wire. The first charge of $e/2$
above half filling results in the appearance of a localized soliton at one of the 
wire edges. Each additional charge of $e/2$ results in an additional delocalized soliton,
so that for charge $Q$ one gets a wire with $2Q-1$ delocalized solitons and a single localized one.

The ground state of the pure Mott state, which is exactly half filled, is obviously insulating.
For wires of odd sizes, the ground states for $Q= \pm 1/2$ are insulating as well,
and the excitation spectrum in these three cases has the regular Mott gap.
Additional insulating states with a non-trivial half gap were found for $Q=\pm 1$.
This unusual gap was explained according to the spectrum of the soliton states in a
1D wire with open boundaries. For $Q > 1$, the excitation spectrum was 
found to be gapless.

The unconventional behavior of wires with $Q > 1$, which are gapless but do not 
preserve translational invariance of charge, is checked against a prediction 
for a TLL wire with an interaction parameter $K<1/2$ coupled to a resonant level. 
We find that the TLL prediction (i.e., the Furusaki-Matveev jump) is indeed obtained also for our system, 
although it involves the creation of two additional delocalized solitons in the charge distribution 
of the wire, as opposed to the inversion of Friedel oscillations in the wire characterizing the 
conventional TLL scenario. As a final remark we note that it may be interesting to 
investigate, for this regime of parameters, some other TLL predictions as well.

\acknowledgments
We thank M. Pepper and M. Kaveh for useful discussions, and
the Israel Science Foundation (Grant 569/07) 
for financial support.

\end{document}